\documentclass[a4paper]{article}
\usepackage[T1]{fontenc}
\usepackage[latin1]{inputenc}
\usepackage[english,francais]{babel}
\usepackage{textcomp}
\usepackage{graphicx}
\usepackage{amsmath}
\usepackage{times}

\begin{document}

\title{Homogeneous cosmology dynamics revealed by Hamiltonian ADM formalism}

\author{Stéphane Fay\footnote{Steph.Fay@Wanadoo.fr} \\
Laboratoire Univers et Théories (LUTH), UMR 8102 \\
Observatoire de Paris, F-92195 Meudon Cedex \\
France}

\date{Composition en \LaTeXe{} le~\today{}}

\maketitle

\selectlanguage{english}

\begin{center}
 \begin{abstract}
We study the homogeneous but anisotropic cosmological models of Bianchi in presence of a massive scalar field $\phi$ using the ADM Hamiltonian formalism. We begin to describe the main steps to find the ADM Hamiltonian of the General Relativity with a massive scalar field and then we study the dynamics of the flat Bianchi type $I$ anisotropic Universe according to initial and final values of this Hamiltonian and sign of the potential. After a brief recall of the conditions necessary to isotropise an anisotropic Bianchi class $A$ model with such a field, we extend them to a non minimally coupled scalar field by using a conformal transformation of the metric which casts the General Relativity with a scalar field into a scalar-tensor theory. The new line element then corresponds to the so-called Brans-Dicke frame, the former one being the Einstein frame. Then, we study the isotropisation process of the Bianchi class $A$ model when we consider the low energy form of the string theory without its antisymmetric tensor and the Brans-Dicke theory with some exponential or power laws of $\phi$ for the potential. Finally, assuming an isotropic Universe such as all the metric functions behave as some power or exponential laws of the proper time, we find the conditions such that the gravitation function and the potential of the scalar field are bounded as it is observed today, and compare them with the necessary conditions for isotropy.
 \end{abstract}
\end{center}

\emph{Please, review the English for style}

\selectlanguage{francais}

\section{Introduction}

\label{s0}

Scalar fields, which  we will refer to by the greek letter $\phi$, are present in many particles physics theories trying to describe the unification of the fundamental forces and to go beyond the interrogations of the standard model. Within this framework, they are interpreted like spin $0$ bosons. Hence the Higgs boson would make it possible to explain the mass of the particles while for supersymmetry theories, they represent additional degrees of freedom necessary to the existence of this symmetry. In this context, it thus appears logical to consider their existence at a cosmological level. Hence, they could be the source of the early times inflation or the recent acceleration of the Universe expansion via what one calls a quintessent scalar field having a negative pressure. They can also play a role in the variation of some constants of the nature like the gravitation constant or the cosmological constant. This one would then become variable and could thus explain the huge difference between its early times predicted value and its value possibly observed today. At a different scale, the scalar fields could also mimic the presence of a dark matter. Hence, in \cite{MatGuzUre99, Fay03A}, it was shown that they could explain the flatness of the rotations curves in the external regions of some galaxies. There are thus many reasons to consider the existence of some scalar fields in the Universe and it is what we will do in this work by studying the General Relativity with a scalar field or a scalar-tensor theory for which $\phi$ is non minimally coupled to the curvature.

From a geometrical point of view, our present Universe is very well described by the homogeneous and isotropic models of Friedmann-Lemaître-Robertson-Walker (FLRW models). Its expansion is then the same everywhere and in any direction. Such symmetry is absolutely extraordinary although it appears natural to us. To make a comparison, it is as extraordinary as a perfect straight line: mathematically it exists but in the real world no straight line is perfectly straight. There are several points of view to explain this. One can admit the isotropy and homogeneity of the Universe as an initial condition. One can suppose the existence of a quantum principle which would select among all the possible geometries, the FLRW models or one can assume that initially our Universe was less symmetrical than today and that it tends asymptotically to an isotropic and homogeneous state. It is this last point of view that we will adopt here by assuming that the Universe is initially homogeneous but anisotropic, which is the most immediate generalisation of the FLRW models. The homogeneous and anisotropic cosmological models were classified by Bianchi at the end of the nineteenth century in nine types listed by Roman letters going from $I$ to $IX$. These types are divided into class $A$ and class $B$, the Bianchi class $A$ models being such as the trace of the structure constants is vanishing.

In this work our goal will be to study the properties of the Bianchi class $A$ models depending on those of the scalar fields with help of the ADM Hamiltonian formalism. In a first section, we give the main steps allowing to calculate the ADM Hamiltonian for the Bianchi class $A$ models in General Relativity with a massive scalar field. In a second section, we show how the dynamics of the metric functions (contraction, expansion, etc) can be deduced from the initial and final values of the Hamiltonian and the sign of the potential for the flat Bianchi type $I$ model. The isotropisation
of the Bianchi class $A$ models in presence of a minimally coupled and massive scalar field was already studied in papers \cite{Fay03, Fay04}. Hence, in a third section, using a conformal transformation of the metric, we show how to generalise these results to a scalar-tensor theory for which the field $\phi$ is non minimally coupled to the curvature and study the cases of the Brans-Dicke theory and low energy string theory without its antisymmetric tensor when their potential are some power or exponential laws of $\phi$. Then, for these two theories, one remarks that when isotropy arises, the metric functions mainly tend to some power or exponential laws of the proper times. This leads us to examine what are the conditions, related to the parameters of these two types of laws, allowing the gravitation function and the potential of the scalar field to be bounded and to compare them with the necessary conditions for isotropisation. One discusses all these results in the last section.

\section{The ADM Hamiltonian formalism}

\label{s122}

There are three main Hamiltonian formulations of General Relativity: they are the Arnowitt, Deser and Misner (ADM) approach, the Dirac and the Kucha\v{r} approaches. The ADM method consists in solving the primary constraints issued from the singular Lagrangian of the theory and then developing the Hamiltonian formulation by using only the independent variables of the phase space. 
We have to remark that in the framework of the gauge theories, this method is far from being ideal: it indeed generally overlooks the covariance with respect to the Poincaré group of symmetry and, in the case of constraints connected to a local invariance gauge, does not always succeed to highlight some aspects of the gauge invariance.
In the present context, the ADM resolution of the constraints greatly simplifies the formalism (which does not include any more superfluous degrees of freedom) and the physical interpretation of the results (it is particularly interesting when one studies the spatial curvature effects on the singularity approach of the anisotropic models).

The Dirac approach is issued from the Dirac constrained systems theory and includes, without solving them, the constraints in the formalism; it is particularly well adapted to the canonical quantification of the theory.
Concerning the Kucha\v{r} formulation, also interesting from the quantification point of view, it enlightens the geometric meaning of the General Relativity Hamiltonian formulation.
In this work, we will follow the ADM method.

The proof of the ADM results is particularly laborious; these calculations are not often explicitly made in the literature. The B appendice of G.~Rossi master thesis (\emph{Formalism hamiltonien
en relativité générale et en cosmologie}, Université de Liège, Faculté des
sciences, Institut de mathématiques, 1973-1974), supervised by J. Demaret, shows the main technical steps. 
Some non published notes of P. Tombal and A. Moussiaux(\emph{Le formalisme
hamiltonien en relativité générale (première version)}, Facultés universitaires Notre-Dame de la Paix, Namur) and a non published document of C. Scheen (\emph{Introduction au formalisme hamiltonien ADM de la relativité générale}, Université de Liège, Institut d'astrophysique et de géophysique, 1992-1993), more systematic and complete, show all the details of calculation. This section is inspired by these three works and proposes a summary of the main technical steps.

The Hamiltonian formalism shows several advantages on the Lagrangian formalism. It allows writing the field equations as a first order equations system instead of the second order. The interpretation of the physical results is easier as shown by the clarity of the singularity chaotic approach of Misner using the ADM formalism with respect to the Belinskii-Khalatnikov-Lifshitz(BKL) one which uses the Lagrangian formalism.

As a first step, we look for the Hamiltonian form of the General Relativity action:
\begin{equation}
\label{fh0}
 S = \int_M R \sqrt{-g} \, d^4x
\end{equation}
\noindent
that we will generalise latter by taking into account a massive and minimally coupled scalar field.

General Relativity is a typical example of a theory having the covariance properties for any coordinate change in spacetime: one says that it is parameterised a priori.
In the classical Hamiltonian theory, it is possible to consider the time variable as a dynamical variable; the parameterisation of this theory reveals some constraints and the variational problem is to "extremalise" a form of the action where these constraints are introduced via Lagrange multipliers.
 Obviously, the constraints do not appear explicitly in the action (\ref{fh0}).
Moreover, in the Hamiltonian formalism, the time, separated from the other variables, is considered as a parameter.
The first thing to do is thus to rewrite the action (\ref{fh0}) by splitting space and time, i.e. by using a $3+1$ spacetime decomposition.
By doing this, we will see that the action can be written in the ADM Hamiltonian form:
\begin{equation}
\label{fh1}
 S
 = \int \biggl[ - g_{ij} \frac{\partial \pi^{ij}}{\partial t}
                - N C_0
                - N^i C_i
                - 2 \Bigl(   \pi^{ij} N_j
                           - \frac{1}{2} N^i \mathrm{Tr} (\pi)
                           + N^{|i} \sqrt{g}
                    \Bigr)_{,i}
        \biggr] \, d^4x
\end{equation}
\noindent
which will allow us to deduce the constraints.

\subsection{Writing the General Relativity action with the $3+1$ spacetime decomposition}

\label{s1221}

\subsubsection{$3+1$ spacetime decomposition}

The $3+1$ spacetime decomposition consists of splitting it as a serie of spatial hypersurfaces, parameterised by the $t$ time. We start to define the \emph{lapse} and \emph{shift} functions.

\begin{figure}[!htbp]
 \begin{center}
  \includegraphics[scale=1.05]{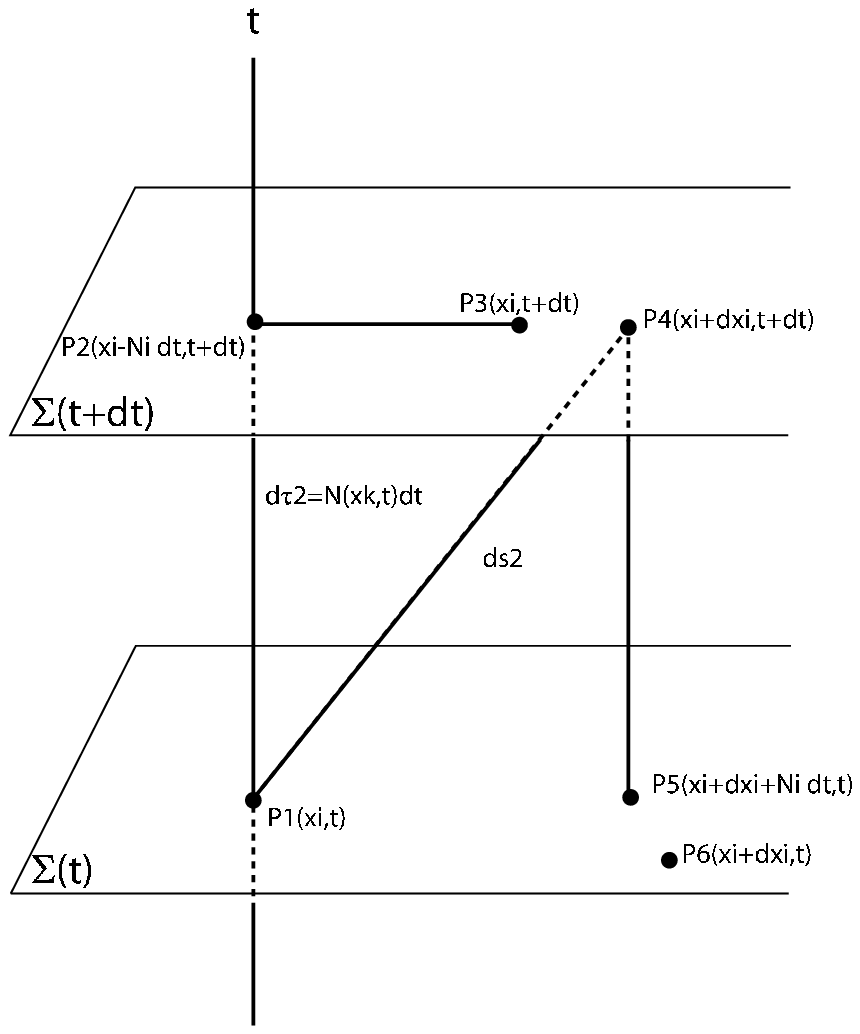}
 \end{center}
\caption{\scriptsize{The $3+1$ spacetime decomposition.}}
\label{fig1}
\end{figure}

Be two hypersurfaces $\Sigma (t)$ and $\Sigma (t + dt)$ represented on the figure \ref{fig1} and their \hbox{3-metrics}, respectively ${}^{(3)}g_{ij} (t, x^k) \, dx^i \, dx^j$ and ${}^{(3)}g_{ij} (t + dt, x^k) \, dx^i \, dx^j$. Be the $P_1$ point with coordinates $(x^i, t)$ on $\Sigma (t)$. We define the $P_2$ point as being the intersection of $\Sigma (t + dt)$ with the normal to $\Sigma (t)$ in $P_1$. The proper time interval $d\tau = N \, dt$ between $P_1$ and $P_2$ then defines the \emph{lapse function} $N (x_k, t)$. Let us define the $P_3$ point on $\Sigma (t + dt)$ as being a point of this hypersurface having the same space coordinates as the $P_1$ point. The $P_3$ point coordinates are thus $(x^i, t + dt)$ whereas these of $P_2$ are $(x^i - N^i \, dt, t + dt)$. The vector binding $P_2$ and $P_3$ then defines the \emph{shift functions} $N^i (x_k, t)$. Be the $P_4$ point on $\Sigma (t + dt)$ with coordinates $(x^i + dx^i, t + dt)$ and the $P_6$ point on $\Sigma (t)$ having the same space coordinates as $P_4$, i.e. $(x^i + dx^i, t)$. We define $P_5$ as being the intersection of the normal to $\Sigma (t + dt)$ in $P_4$ with $\Sigma (t)$. The $P_5$ coordinates are then $(x^i + dx^i + N^i \, dt, t)$.

We are now able to express the line element $ds^2$ between the $P_1$ and $P_4$ points with help of the \hbox{3-metric}~${}^{(3)}g_{ij}$, the \emph{shift} and the \emph{lapse} functions. Writing the Pythagoras theorem in the non Euclidean \hbox{4-space} with signature $({-}, {+}, {+}, {+})$, it comes:
\begin{align*}
 ds^2
 & =   {}^{(4)}g_{\alpha \beta} \, dx^\alpha \, dx^\beta \\
 & =   {}^{(3)}g_{ij} (x^k, t) \bigl( x^i (P_5) - x^i (P_1) \bigr)
                               \bigl( x^j (P_5) - x^j (P_1) \bigr)
     - d \tau^2 \\
 & =   {}^{(3)}g_{ij} (x^k, t) (dx^i + N^i \, dt) (dx^j + N^j \, dt)
     - N^2 \, dt^2
\end{align*}
\noindent
from which we get for the metric:
\begin{displaymath}
 {}^{(4)}g_{\alpha \beta}
 = \begin{pmatrix}
    {}^{(4)}g_{00} & {}^{(4)}g_{0j} \\
    {}^{(4)}g_{i0} & {}^{(4)}g_{ij}
   \end{pmatrix}
 = \begin{pmatrix}
    - N^2 + {}^{(3)}g_{ij} N^i N^j & {}^{(3)}g_{ij} N^i \\
            {}^{(3)}g_{ij} N^j     & {}^{(3)}g_{ij}
   \end{pmatrix}
\end{displaymath}
\noindent
or, putting $N_i \doteq {}^{(3)}g_{ij} N^j$~:
\begin{equation}
\label{ea1}
 {}^{(4)}g_{\alpha \beta}
 = \begin{pmatrix}
    N_k N^k - N^2 & N_j            \\
    N_i           & {}^{(3)}g_{ij}
   \end{pmatrix}
\end{equation}
\noindent
and using ${}^{(4)}g_{\alpha \beta} {}^{(4)}g^{\beta \gamma}
= \delta_\alpha^\gamma$~:
\begin{equation}
 {}^{(4)}g^{\alpha \beta}
 = \begin{pmatrix}
    - N^{-2}     & N_j N^{-2}                      \\
      N_i N^{-2} & {}^{(3)}g_{ij} - N^i N^j N^{-2}
   \end{pmatrix}
\end{equation}
\indent
To calculate the determinant ${}^{(4)}g$ of the \hbox{4-metric}, we use the Frobenius-Schur theorem which shows that if $A$, $B$, $C$ and $D$ are four square matrices, the matrix determinant:
\begin{equation}
 \Delta
 \doteq \begin{pmatrix}
         A & B \\
         C & D
        \end{pmatrix}
\end{equation}
\noindent
is $\det (\Delta) = \det (D) \det (A - B D^{-1} C)$.
We then deduce:
\begin{equation}
\label{ea2}
 \sqrt{-{}^{(4)}g}
 = N \sqrt{{}^{(3)}g}
\end{equation}
\indent
Before continuing, we have to define the concept of \emph{extrinsic curvature}. The extrinsic curvature characterises the way in which a variety is included in a space of higher dimension. For example, a paper sheet with two dimensions which one twists in a space with three dimensions has an extrinsic curvature with respect to this space.
On the contrary, our Universe has an \emph{intrinsic curvature} which does not need some higher dimension to be defined. In the case which interests us, the extrinsic curvature of a spatial hypersurface is a measure of the direction variation of the normal $\vec{n}$ to the hypersurface $\Sigma (t)$ between two infinitely close points on $\Sigma(t)$ and is defined by:
\begin{equation}
\label{ea3}
 K_{ij}
 \doteq - n_{i;j}
 =      - \vec{e}_j \cdot \nabla_i \vec{n}
\end{equation}

\subsubsection{The Gauss-Weingarten and Gauss-Codazzi relations:
rewriting the action}
When the extrinsic curvature is known, one can express the covariant derivative of the basis vectors $\vec{e}_j$ of the spatial hypersurface $\Sigma$ in the spacetime, depending on quantities only related to $\Sigma$. These are the Gauss-Weingarten relations which write:
\begin{equation}
\label{ea4}
 {}^{(4)}\nabla_i \vec{e}_j
 = - K_{ij} \vec{n} + {}^{(3)}\Gamma^k_{ij} \vec{e}_k
\end{equation}
\indent
The Gauss-Codazzi relations try to express the spacetime intrinsic curvature depending on the intrinsic and extrinsic curvatures of the hypersurface. They express as:
\begin{align}
 {}^{(4)}R^0_{ijk}
 & = K_{ik|j} - K_{ij|k}
 \label{ea5} \\
 {}^{(4)}R_{mijk}
 & = - (K_{ij} K_{mk} - K_{ik} K_{mj}) + {}^{(3)}R_{mijk}
 \label{ea6} \\
 {}^{(4)}R_{i0k0}
 & = K_{ik, n} + K^m_k K_{im}
 \label{ea7}
\end{align}
\noindent
where $|$ stands for the covariant derivative on the hypersurface. These relations allow rewriting the scalar curvature depending on the intrinsic and extrinsic curvatures on the hypersurface. Indeed, one can rewrite:
\begin{displaymath}
 {}^{(4)}R
 = {}^{(4)}R^{ij}_{ij} - 2 {}^{(4)}R^j_{0j0}
\end{displaymath}
\indent
The right hand side of this equation only contains some terms expressed by the Gauss-Codazzi relations and if we define $\mathrm{Tr} (K^2) \doteq K^{jk} K_{jk}$ and $K \doteq K^i_i \doteq \mathrm{Tr} (K)$, one gets:
\begin{displaymath}
 {}^{(4)}R^{ij}_{ij}
 = g^{ik} g^{jm} {}^{(4)}R_{kmij}
 = {}^{(3)}R - \mathrm{Tr} (K^2) + K^2
\end{displaymath}
\noindent
and:
\begin{displaymath}
 {}^{(4)}R^j_{0j0}
 = g^{jk} {}^{(4)}R_{k0j0}
 = K_{,n} + \mathrm{Tr} (K^2) - K_{kj} g^{jk}_{,n}
\end{displaymath}
\indent
Since moreover, one can show that:
\begin{displaymath}
 g^{jk}_{,n}
 = 2 K^{jk}
\end{displaymath}
\noindent
it comes for the scalar curvature:
\begin{equation}
\label{ea8R}
 {}^{(4)}R
 = {}^{(3)}R + \mathrm{Tr} (K^2) + K^2 - 2 K_{,n}
\end{equation}
\indent
Using this last relation and (\ref{ea2}), we rewrite the Hilbert action (\ref{fh0}) in the following way:
\begin{equation}
\label{ea8S}
 S
 =   \int_M N \sqrt{{}^{(3)}g}
     \bigl( {}^{(3)}R + \mathrm{Tr} (K^2) - K^2 \bigr) \, d^4x
   - 2 N \int_{\partial M} K \sqrt{{}^{(3)}g} \, d^3x
\end{equation}
\indent
The surface term may be removed by imposing specific conditions on the variety border or by adding to the initial action another surface term compensating the one of (\ref{ea8S}) -- in this last case, we remove the surface term by taking as the initial action:
\begin{equation}
\label{ea9}
 S
 =   \int_M {}^{(4)}R \sqrt{- {}^{(4)}g} \, d^4x
   + 2 \int_{\partial M} K \sqrt{{}^{(3)}g} \, d^3x
\end{equation}
\indent
In the case of a closed space, removing the surface term is not a problem: there is no consequence on the field equations when one varies the geometry inside the surface border of the variety. However, for open spaces which are asymptotically flat, it is necessary to add a surface term.

Whatever the way in which one gets rid of the surface term, one obtains:
\begin{equation}
\label{ea8a}
 S
 = \int_M N \sqrt{{}^{(3)}g}
   \bigl( {}^{(3)}R + \mathrm{Tr} (K^2) - K^2 \bigr) \, d^4x
\end{equation}
\indent
In what follows, we are going to rewrite the action (\ref{ea9}) with help of the $3+1$ decomposition

\subsubsection{Writing the action as a $3+1$ decomposition}

To write the above action as a $3+1$ decomposition, we have to make this transformation for:
\begin{enumerate}
 \item the Christoffel symbols which are:
 \begin{displaymath}
  \Gamma_{\alpha \beta}^\gamma
  = \frac{1}{2} g^{\gamma \delta}
    \bigl(   g_{\alpha \delta, \beta}
          + g_{\delta \beta, \alpha}
          - g_{\alpha \beta, \delta}
    \bigr)
 \end{displaymath}
 \item the Ricci tensor which writes:
 \begin{displaymath}
  {}^{(4)}R_{ij}
  =   \Gamma^\alpha_{ij, \alpha}
    - \Gamma^\alpha_{i \alpha, j}
    + \Gamma^\alpha_{ij} \Gamma^\beta_{\alpha \beta}
    - \Gamma^\alpha_{i \beta} \Gamma^\beta_{j \alpha}
 \end{displaymath}
 \item the scalar curvature which is ${}^{(4)}R = {}^{(4)}R^\alpha_\alpha$, and thus the action.\\
\end{enumerate}

To rewrite the Christoffel symbols $\Gamma$, we introduce the following definition:
\begin{displaymath}
 \xi^j
 \doteq \frac{N^j}{N}
\end{displaymath}
\noindent
and define the components of the tensor $\Lambda$, the Christoffel \hbox{3-symbols} relative to the hypersurface, as:
\begin{displaymath}
 \Lambda_{ik}^j
 \doteq \frac{1}{2} {}^{(3)}g^{jm}
        \bigl( {}^{(3)}g_{im, k} + {}^{(3)}g_{mk, i} - {}^{(3)}g_{ik, m}
        \bigr)
\end{displaymath}
\indent
After tedious calculations, we get the following forms for the Christoffel symbols:
\begin{align*}
 \Gamma_{00}^0
 & = \frac{1}{N} \partial_0 N + N_{|i} \xi^i - N \xi^i \xi^j K_{ij} \\
 \Gamma_{00}^i
 & =   N \gamma^{ij} \partial_0 \xi_j
     + \frac{1}{2} \gamma^{ij} \bigl( N^2 (1 - \xi_m \xi^m) \bigr)_{,j}
     - N N_{|j} \xi^i \xi^j
     + N^2 \xi^i \xi^j \xi^k K_{jk} \\
 \Gamma_{ik}^j
 & = \Lambda^j_{ik} + \xi^j K_{ik} \\
 \Gamma_{0i}^0
 & = \frac{N_{|i}}{N} - K_{ij} \xi^j \\
 \Gamma_{i0}^j
 & = N (- K^j_i + \xi^j_{|i} + \xi^j K_{im} \xi^m)
\end{align*}
\indent
We inject them in the Ricci tensor spatial components  ${}^{(4)}R_{ij}$ which express depending on the Christoffel symbols as:
\begin{displaymath}
 {}^{(4)}R_{ij}
 =   \Gamma^\alpha_{ij, \alpha}
   - \Gamma^\alpha_{i \alpha, j}
   + \Gamma^\alpha_{ij} \Gamma^\beta_{\alpha \beta}
   - \Gamma^\alpha_{i \beta} \Gamma^\beta_{j \alpha}
\end{displaymath}
\indent
Then, it comes:
\begin{align}
 {}^{(4)}R_{ij}
 & =   {}^{(3)}R_{ij}
     - \frac{1}{N} \partial_0 K_{ij}
     - \frac{1}{N} \bigl( N_{|ij} - N_{|i} K_{jk} \xi^k - N_{|j} K_{ik} \xi^k
                   \bigr)
 \nonumber \\
 & \qquad
     + K_{ij} K
     - 2 K_{ik} K^k_j
     + K_{ik} \xi^k_{|j}
     + K_{jk} \xi^k_{|i}
     + \xi^k K_{ij|k}
 \label{ricci-ij}
\end{align}
\noindent
where ${}^{(3)}R_{ij}$ is the Ricci tensor of a hypersurface and is thus conventionally defined as a function of its Christoffel symbols as:
\begin{displaymath}
 {}^{(3)}R_{ij}
 =   \Lambda^k_{ij, k}
   - \Lambda^k_{ik, j}
   + \Lambda^k_{ij} \Lambda^l_{kl}
   - \Lambda^k_{il} \Lambda^l_{jk}
\end{displaymath}
\indent
To continue our calculation, we should also rewrite explicitly the ${}^{(4)}R_{0i}$ and ${}^{(4)}R_{00}$ components of the Ricci tensor.
However, these calculations are definitely harder than those which lead to the purely spatial components (\ref{ricci-ij}) of the Ricci tensor. In fact in General Relativity, it is enough to use a reference frame which simplifies the calculation without occulting the information related to the reference frame degrees of freedom. We will use the reference frame defined by the relations:
\begin{align*}
 \vec{n}
 & \doteq   \frac{1  }{N} \frac{\partial}{\partial t}
          - \frac{N^i}{N} \frac{\partial}{\partial x^i} \\
 \vec{e}_i
 & \doteq   \frac{\partial}{\partial x^i}
\end{align*}
\indent
In this particular frame, we have $g_{nn} = \vec{n} \cdot \vec{n} = -1$, $g_{ni} = \vec{n} \cdot \vec{e}_i = 0$, $g_{ij} = \vec{e}_i \cdot \vec{e}_j$. The scalar curvature ${}^{(4)}R$ writes then as ${}^{(4)}R = 2 G_{n}^{n} + 2 g^{ij} {}^{(4)}R_{ij}$, where $G_{\alpha \beta}$ is the Einstein tensor. 
In virtue of the Gauss-Codazzi relations (\ref{ea6}), we get:
\begin{displaymath}
 G_n^n
 = - \frac{1}{2} \bigl( {}^{(3)}R + K^2 - K_{ij} K^{ij} \bigr)
\end{displaymath}
\indent
We introduce this expression and (\ref{ricci-ij}) in the scalar curvature ${}^{(4)}R$ and get:
\begin{align*}
 {}^{(4)}R
 & =   {}^{(3)}R
     + K^2
     - K_{ij} K^{ij}
     - \frac{2}{N} g^{ij} \partial_0 K_{ij}
     - \frac{2}{N} N^{|i}_{|i}
     + \frac{4}{N} N_{|i} K^i_k \xi^k
 \\
 & \qquad
     - 2 K_{ij} K^{ij}
     + 4 K^i_k \xi^k_{|i}
     + 2 g^{ij} K_{ij|k} \xi^k
\end{align*}
\noindent
where ${}^{(3)}R$ is the scalar curvature relative to the hypersurface; it expresses as:
\begin{displaymath}
 {}^{(3)}R
 = 2 {}^{(3)}R^{12}_{12} + 2 {}^{(3)}R^{13}_{13} + 2 {}^{(3)}R^{23}_{23}
\end{displaymath}
\indent
Finally, the action in the $3+1$ decomposition will take the following form:
\begin{align}
 S
 & = \int_M d^4x \, \sqrt{g} \,
     \bigg[ - g^{ij} \frac{\partial K_{ij}}{\partial t}
            - \frac{\partial K}{\partial t}
            + N \bigl( {}^{(3)}R + K^2 - K_{ij} K^{ij} \bigr)
            + 2 N^i \delta^j_i K_{|j}
 \nonumber \\
 & \qquad \qquad \qquad \qquad
            - 2 N^{|i}_{|i}
            + 2 N^{|i} K_{ij} \xi^j
            + 2 N K_{ij} \xi^{j|i}
     \bigg]
 \label{ea10}
\end{align}

\subsection{Identification of the action $3+1$ form with its form in the Hamiltonian formalism}

\label{s1222}

In this section, we wish to show the equivalence between the form (\ref{ea10}) of the action and its form adapted to the Hamiltonian formalism:
\begin{equation}
\label{id0}
 S
 = \int_M d^4x \,
   \biggl[ - g_{ij} \frac{\partial \pi^{ij}}{\partial t}
           - N C_0
           - N^i C_i
           - 2 \Bigl(   \pi^{ij} N_j
                      - \frac{1}{2} N^i \mathrm{Tr} (\pi)
                      + N^{|i} \sqrt{g}
               \Bigr)_{,i}
   \biggr]
\end{equation}
\indent
For that, we define the canonically conjugate momenta $\pi^{ij}$:
\begin{equation}
\label{id1}
 \pi^{ij}
 \doteq \sqrt{g} (g^{ij} K - K^{ij})
\end{equation}
\noindent
and introduce the superhamiltonian as:
\begin{equation}
\label{id2}
 C_0
 = - ({}^{(3)}R + K^2 - K_{ij} K^{ij}) \, \sqrt{g}
 = - \sqrt{g} {}^{(3)}R + \sqrt{g} (K_{ij} K^{ij} - K^2)
\end{equation}
\indent
However, we calculate that:
\begin{displaymath}
 K_{ij} K^{ij} - K^2
 = \frac{1}{g} \biggl[   \mathrm{Tr} (\pi^2)
                       - \frac{1}{2} \bigl( \mathrm{Tr} (\pi) \bigr)^2
               \biggr]
\end{displaymath}
\noindent
and consequently, it comes:
\begin{equation}
\label{id3}
 C_0
 = - \sqrt{g} {}^{(3)}R
   + \frac{1}{\sqrt{g}} \biggl[   \mathrm{Tr} (\pi^2)
                                - \frac{1}{2} \bigl( \mathrm{Tr} (\pi) \bigr)^2
                        \biggr]
\end{equation}
\indent
In addition, we can also show that:
\begin{equation}
\label{id4}
 - g_{ij} \frac{\partial \pi^{ij}}{\partial t}
 = \sqrt{g} \biggl( - g^{ij} \frac{\partial K_{ij}}{\partial t}
                    - \frac{\partial K}{\partial t}
            \biggr)
\end{equation}
\indent
Last, we introduce the supermomenta $C_i$ \emph{via}~:
\begin{equation}
\label{id5}
 - N_i C^i
 = - 2 N^i (K_i^j - \delta_i^j K)_{|j} \sqrt{g}
 =   2 N^i \bigl[ - \sqrt{g} (K_i^j - \delta_i^j K) \bigr]_{|j}
\end{equation}
\noindent
that we can rewrite:
\begin{align*}
 C_i
 & = - 2 \pi^j_{i|j}
   =   2 \sqrt{g} (K^j_i - \delta^j_i K)_{|j} \\
 C^i
 & = - 2 \pi^{ij}_{|j}
   =   2 \sqrt{g} (K^{ij} - g^{ij} K)_{|j}
\end{align*}
\indent
By inserting all these results in the action (\ref{id0}), we recover, after some additional calculation, the action written with the $3+1$ decomposition.

\subsection{Formulation of the ADM constraints in the General Relativity}

\label{s1223}

The canonical momenta $\pi^{ij}$ are naturally defined by:
\begin{displaymath}
 \pi^{ij}
 \doteq \frac{\delta L}{\delta \dot{g}_{ij}}
\end{displaymath}
\noindent
$L$ being the Lagrangian.
The action of the Hamiltonian formalism writes:
\begin{equation}
\label{fa0}
 S
 = \int_M \Bigl( \pi^{ij} \frac{\partial g_{ij}}{\partial t} - N C_0 - N^i C_i
          \Bigr) \, d^4x
\end{equation}
\indent
The demonstration of the equivalence between the above expression and the form (\ref{fh0}) of the action is similar to assure the equivalence between the actions (\ref{ea8a}) and (\ref{fh1}). Consequently, by varying (\ref{fa0}) with respect to the \emph{lapse} and \emph{shift} functions, considered as some Lagrange multipliers, we get the constraints:
\begin{align}
 C_0
 & = - \sqrt{g} {}^{(3)}R
     + \frac{1}{\sqrt{g}}
       \biggl[   \mathrm{Tr} (\pi^2)
               - \frac{1}{2} \bigl( \mathrm{Tr} (\pi) \bigr)^2
       \biggr]
   =   0
 \label{c0} \\
 C_i
 & = - 2 \pi_{i|j}^j
   =   0
 \label{con1}
\end{align}
\indent
The constraints rule the dynamics of the geometry and are, in the same time, some initial values conditions. Because of the constraints $C_\mu = 0$, it is possible to choose freely the fields ${}^{(3)}g_{ij}$ and $\pi^{ij}$ on the initial hypersurface $\Sigma(t_0)$.
The dynamical equations rule the changes of the intrinsic geometry and the extrinsic curvature of a hypersurface when one moves from a hypersurface to a neighbour hypersurface.

If the constraints are satisfied on $\Sigma (t_0)$ and if the canonically conjugate fields evolve and check the dynamical equations, then the constraints are conserved all the time.

In virtue of the constraints, four of the ${}^{(3)}g_{ij}$ and $\pi^{ij}$ fields may be expressed with the others; moreover, imposing the coordinate conditions fixes four of the
remaining fields. Only remain two couples of canonical variables; the following paragraph shows how the ADM approach is brought back to these two physical degrees of freedom.

\subsection{ADM formulation of the General Relativity with a minimally coupled and massive scalar field}

\label{s1224}

To know how are modified the constraints equations (\ref{c0}) and (\ref{con1}) when a scalar field $\phi$ is present, we are going to write the $3+1$ decomposition of the part of the action with the scalar field:
\begin{equation}\label{actionE}
 S
 = \int \biggl[   R
                - \frac{1}{2} \frac{2 \omega + 3}{\phi^2}
                  \phi_{,\mu} \phi^{,\mu}
                - U
        \biggr] \, \sqrt{-{}^{(4)}g} \, d^4x
\end{equation}
\indent
where $\omega(\phi)$ and $U(\phi)$ are respectively the Brans-Dicke coupling function describing the coupling between the metric and the scalar field and the potential of the scalar field describing the self coupling of $\phi$. First, we rewrite the term containing $\omega$ in the following form:
\begin{equation}
\label{fst1}
 - (3/2 + \omega) \phi^{,\mu} \phi_{,\mu} \phi^{-2} \sqrt{-{}^{(4)}g}
 = (3/2 + \omega) \dot{\phi}^2 \phi^{-2} N^{-1} \sqrt{g}
\end{equation}
\indent
Taking into account this last expression, we can express the scalar field conjugate momentum as:
\begin{equation}
\label{fst2}
 \pi_\phi
 \doteq \frac{\partial I}{\partial \dot{\phi}}
 =     (3 + 2 \omega) \dot{\phi} \phi^{-2} N^{-1} \sqrt{g}
\end{equation}
\indent
where $I$ is the Lagrangian of the above action. From this last equation, we deduce:
\begin{equation}
\label{fst3}
 \dot{\phi}
 = \pi_\phi \frac{N}{\sqrt{g}} \frac{\phi^2}{3 + 2 \omega}
\end{equation}
\noindent
which allows us to rewrite (\ref{fst1}) in the following way:
\begin{align*}
 - (3/2 + \omega) \phi^{,\mu} \phi_{,\mu} \phi^2 \sqrt{{}^{(4)}g}
 & = \frac{(3/2 + \omega)}{\phi^2} \frac{1}{N} \sqrt{g} \pi_\phi^2
     \frac{N^2}{g} \frac{\phi^4}{(3 + 2 \omega)^2} \\
 & = \frac{1}{2} \frac{\phi^2}{3 + 2 \omega} \frac{N}{\sqrt{g}} \pi_\phi^2
\end{align*}
\indent
To the constraint (\ref{c0}), we have thus to add the terms $C_{0\phi}$ coming from the Brans-Dicke coupling function and the potential, $C_{0\phi}$ being such as:
\begin{equation}
   \frac{1}{2} \frac{\phi^2}{3 + 2 \omega} \frac{N}{\sqrt{g}} \pi_\phi^2
 - N U \sqrt{g}
 = \pi_\phi \dot{\phi} - N C_{0\phi}
\end{equation}
\indent
Thus, using the expression for $\pi_\phi$ given by (\ref{fst2}) and of $\dot{\phi}$, it comes:
\begin{align*}
 C_{0\phi}
 & = - \frac{1}{2} \frac{\phi^2}{3 + 2 \omega} \frac{1}{\sqrt{g}} \pi_\phi^2
     + \pi_\phi \frac{\dot{\phi}}{N}
     + U \sqrt{g} \\
 & = - \frac{1}{2} \frac{\phi^2}{3 + 2 \omega} \frac{1}{\sqrt{g}} \pi_\phi^2
     + \frac{\phi^2}{3 + 2 \omega} \frac{1}{\sqrt{g}} \pi_\phi^2
     + U \sqrt{g} \\
 & =   \frac{1}{2} \frac{\phi^2}{3 + 2 \omega} \frac{1}{\sqrt{g}} \pi_\phi^2
     + U \sqrt{g}
\end{align*}
\indent
The final form of the $C_0$ constraint, taking into account the scalar field, is thus:
\begin{displaymath}
 C_0
 = - \sqrt{g} {}^{(3)}R
   + \frac{1}{\sqrt{g}}
     \biggl[   \mathrm{Tr} (\pi^2)
             - \frac{1}{2} \bigl( \mathrm{Tr} (\pi) \bigr)^2
     \biggr]
   + \frac{1}{2} \frac{\phi^2}{3 + 2 \omega} \frac{1}{\sqrt{g}} \pi_\phi^2
   + U \sqrt{g}
\end{displaymath}
\indent
The general form of the diagonal homogeneous but anisotropic metrics of Bianchi is defined by:
\begin{equation}\label{metric}
 ds^2
 = - N^2 d\Omega
   + R_0^2 e^{- 2 \Omega}
     \Bigl(   e^{\beta_+ + \sqrt{3} \beta_-} (\omega^1)^2
            + e^{\beta_+ - \sqrt{3} \beta_-} (\omega^2)^2
            + e^{- 2 \beta_+}                (\omega^3)^2 
     \Bigr)
\end{equation}
\noindent
where the $\omega_i$ are the 1-forms specifying each Bianchi model. This metric is anisotropic because it describes a Universe where the expansion rate depends on the direction. The Misner parameterisation \cite{Mis69,Mis69A} of the conjugate momenta expresses as:
\begin{align*}
 p_k^i
 & = 2 \pi \pi_k^i - \frac{2}{3} \pi \delta_k^i \pi_l^l \\
 6 p_{ij}
 & = \mathrm{diag} \bigl( p_+ + \sqrt{3} p_-, p_+ - \sqrt{3} p_-, - 2 p_+
                   \bigr)
\end{align*}
\noindent
Then, we can rewrite the constraint on the following form:
\begin{align*}
 C_0
 & = - R_0^3 e^{- 3 \Omega}
       \biggl[   {}^{(3)}R
               + \frac{1}{R_0^6 e^{- 6 \Omega}}
                 \biggl(   \frac{1}{6} (\pi_k^k)^2
                         - \frac{1}{24 \pi^2} (p_+^2 + p_-^2)
                 \biggr)
       \biggr]
 \nonumber \\
 & \qquad
     + \frac{1}{2 R_0^3 e^{- 3 \Omega}}
       \frac{\phi^2 \pi_\phi^2}{(3 + 2 \omega)}
     + R_0^3 e^{- 3 \Omega} U
\end{align*}
\noindent
$\pi$ being the number in this last expression, and the action becomes:
\begin{displaymath}
 S
 = \int p_+ \, d\beta_+ + p_- \, d\beta_- + p_\phi \, d\phi - H \, d\Omega
\end{displaymath}
\noindent
with the ADM Hamiltonian $H = 2 \pi \pi^k_k$ and the scalar field conjugate momentum $p_\phi = \pi \Pi_\phi$.
By using the constraint $C_0 = 0$, we then deduce for $H$:
\begin{equation}
\label{fstHam}
 H^2
 =   p_+^2
   + p_-^2
   + 12 \frac{p_\phi^2 \phi^2}{3 + 2 \omega}
   + 24 \pi^2 R_0^6 e^{- 6 \Omega} U+V(\Omega,\beta_+,\beta_-)
\end{equation}
\indent
where $V(\Omega,\beta_+,\beta_-)$ are the curvature terms specified in the table \ref{tabCurv} for each Bianchi class $A$ model. For the class $B$, the ADM Hamiltonian formulation  does not give the correct field equations as explained in \cite{RyaWal84} and we have to redefine the divergence theorem in a non-coordinated basis to solve this problem and get a correct variational principle. The physical degrees of freedom have been isolated, but with this form the theory is no more covariant -- the constraints have been solved and the coordinate conditions have been fixed. The loss of covariance is obvious besides: the ADM Hamiltonian is not null, while the cancellation of the Hamiltonian is characteristic of the constrained systems.
\begin{table}[h]
\begin{center}
\begin{tabular}{|l|l|}
\hline
Bianchi type&$V(\Omega,\beta_+,\beta_-)$\\
\hline
$I$&0\\
\hline
$II$&$12\pi^2R_0^4e^{4(-\Omega+\beta_++\sqrt{3}\beta_-)}$\\
\hline
$VI_0$, $VII_0$&$24\pi^2R_0^4e^{-4\Omega+4\beta_+}(\cosh{4\sqrt{3}\beta_-}\pm 1)$\\
\hline
$VIII$, $IX$&$24\pi^2R_0^4e^{-4\Omega}\mbox{[}e^{4\beta_+}(\cosh{4\sqrt{3}\beta_-}-1)+$\\
&$1/2e^{-8\beta_+}\pm2e^{-2\beta_+}\cosh{2\sqrt{3}\beta_-}\mbox{]}$\\
\hline
\end{tabular}
\caption{\label{tabCurv}\scriptsize{Form of the curvature potential $V(\Omega,\beta_+,\beta_-)$ for each Bianchi model.}}
\end{center}
\end{table}

\section{Dynamical study of the General Relativity with a massive scalar field for the flat Bianchi type $I$ model}

\label{s2}

The results of this section are issued from papers \cite{Fay00A} published in Classical and Quantum Gravity copyright 2000 IOP Publishing Ltd. We aim to describe the global properties of the metric functions (expansion, contraction and extrema) depending on Hamiltonian initial and final values and the sign of the potential when we consider the General Relativity with a massive scalar field in a flat Bianchi type $I$ model, the corresponding Hamiltonian and metric being respectively defined by (\ref{fstHam}) and (\ref{metric}) with $(\omega_1,\omega_2,\omega_3)=(dx,dy,dz)$. From \cite{Nar72}\footnote{(see \cite{Nar72}, p1830 for a detailed calculus)}, we know that the lapse function $N$ takes the form:
\begin{equation} \label{18} 
N=\frac{12\pi R_0 ^3 e^{-3\Omega}}{H} 
\end{equation}
and then using the relation $dt=-N d\Omega$, it comes:

\begin{equation} \label{20} 
\frac{dg_{ij}}{dt}=2R_0^2 (\frac{d\beta_{ij}}{dt}-\frac{d\Omega}{dt})e^{-2\Omega+2\beta_{ij}}= 2R_0 ^2 e^{-2\Omega+2\beta_{ij}}\frac{H-p_{ij}}{HN} 
\end{equation}
To obtain the global properties of the metric functions, we have to study the sign of the quantity (\ref{20}) which depends on $H-p_{ij}$. For sake of simplicity, we will consider a potential with a constant sign. We will see later how to extend our results when the sign of the potential varies. For the Bianchi type $I$ model, the Hamiltonian equation
\begin{equation} \label{17} 
\dot{H}=\frac{dH}{d\Omega}=\frac{\partial H}{\partial \Omega}=-72\pi^2 R_0 ^6 \frac{e^{-6\Omega}U}{H} 
\end{equation}
where the dot means a derivative with respect to $\Omega$, shows that $\dot H$ and $H$ are monotonic functions of $\Omega$ with constant sign. Then, from the lapse function, it comes that $\Omega$ is a monotonic function of the proper time $t$ and so for $H(t)$. If $t$ varies from $t_{ini}$ to $t_{end}$, H will thus evolve monotonically from $H(t_{ini})=H_{ini}$ to $H(t_{end})=H_{end}$.\\ Moreover, from the Hamiltonian equations 
\begin{equation} \label{15} 
\dot{p}_\pm=-\frac{\partial H}{\partial \beta_ \pm}=0 
\end{equation}
we derive that the conjugate momenta $p_{ij}$ are some constants (this is not the case in presence of curvature). Hence, the equation (\ref{20}) may only have one zero and \emph{when the potential has a constant sign on a proper time interval, there is to the more one single extremum for the metric functions}.\\
\\
From (\ref{20}), it is clear that the metric function derivative will vanish if the three conditions $C_1$, $C_2$ and $C_3$ are true:
\begin{itemize}
\item $C_1$: H and $p_{ij}$ have the same sign
\item $C_2$ and $C_3$: $p_{ij}$ belongs to the interval defined by $H_{ini}$ and $H_{end}$
\end{itemize}
Hence, we have to consider the following four cases for which we give the variation of the metric function with respect to the proper time $t$:
\newline
\newline
\emph{Case 1a: $U<0$, $\dot{H}$ and $H>0$}
\newline
Taking into account (\ref{18}) and the relation $dt=-Nd\Omega$, it comes that when $\Omega$ increases, $t$ decreases. The Hamiltonian is thus a decreasing function of $t$ and the three conditions $C_i$ write:
\begin{itemize}
\item $C_1$: $p_{ij}>0$
\item $C_2$: $H_{fin}-p_{ij}$<0
\item $C_3$: $H_{ini}-p_{ij}$>0
\end{itemize}
Whatever the case we consider, if $C_2$ or $C_3$ are false, respectively $C_3$ or $C_2$ are true. In addition, in the present case, if $C_3$ is false, $C_1$ is true. 

If the three conditions are true, the metric function has a maximum in the proper time of the Einstein frame since the Hamiltonian will be equal to $p_{ij}$ for a value of $\Omega$.

If $C_1$ is wrong, it is increasing since then $H-p_{ij}$ has always the sign of $H$.

If $C_2$ is wrong, the Hamiltonian is always larger than $p_{ij}$, and the metric function is again increasing for the $t$ time.

If $C_3$ is wrong, $C_1$ is true, and the metric function decreases since the Hamiltonian is always smaller than $p_{ij}$.
\newline
\newline
The same reasoning will hold for the other cases.
\newline
\newline
\emph{Case 1b: $U<0$, $\dot{H}$ and $H<0$}
\newline
When $\Omega$ increases, $t$ is increasing. The Hamiltonian is a negative and decreasing function of these times coordinates. When $C_2$ is wrong, $C_1$ is true. The three conditions can be written as:
\begin{itemize}
\item $C_1$: $p_{ij}<0$
\item $C_2$: $H_{fin}-p_{ij}$<0
\item $C_3$: $H_{ini}-p_{ij}$>0
\end{itemize}

If they are all true, the metric function has a maximum.

If $C_1$ or $C_3$ is false, it is decreasing.

If $C_1$ is true, $C_2$ is false and it is increasing.
\newline
\newline
\emph{Case 2a: $U>0$, $\dot{H}<0$ and $H>$0}
\newline
When $\Omega$ increases, $t$ decreases. The Hamiltonian is a positive and decreasing function of $\Omega$ and then an increasing function of $t$. When $C_2$ is wrong, $C_1$ is true. The three conditions can be written as:
\begin{itemize}
\item $C_1$: $p_{ij}>0$
\item $C_2$: $H_{end}-p_{ij}$>0
\item $C_3$: $H_{ini}-p_{ij}$<0
\end{itemize}

If they are all true, the metric function has a minimum.

If $C_1$ or $C_3$ is false, it is increasing.

If $C_2$ is false, $C_1$ is true, and the metric function is decreasing.
\newline
\newline
\emph{Case 2b: $U>0$, $\dot{H}>0$ and $H<$0}
\newline
When $\Omega$ increases, $t$ increases. The Hamiltonian is a negative and increasing function of the two time coordinates. When $C_3$  is false, $C_1$ is true. We obtain for the three conditions:
\begin{itemize}
\item $C_1$: $p_{ij}<0$
\item $C_2$: $H_{end}-p_{ij}$>0
\item $C_3$: $H_{ini}-p_{ij}$<0
\end{itemize}

If the three conditions are true, the metric function has a minimum.

If $C_1$ or $C_2$ is false, it is decreasing.

If $C_1$ is true, $C_3$ is false, the metric function is increasing.
\newline
\newline
All these results are summarised in table \ref{tab1}. They can not be extended to the other Bianchi models with curvature. They will be commented in the discussion. In the next section, we study the isotropisation of all the Bianchi class $A$ models for a non minimally coupled and massive scalar field.
\begin{table}[h]
\begin{tabular}{@{}lllll}
\hline
 & $H$, $\dot{H}>0$,& $H$, $\dot{H}$,& $H$, $U>0$& $\dot{H}$, $U>0$,\\
&$U<0$&$U<0$&$\dot{H}<0$&$H<0$\\
\hline
$C_1$, $C_2$, $C_3$: true & Maximum & Maximum & Minimum & Minimum 		\\
$C_1$: false 		& Increasing & Decreasing & Increasing & Decreasing \\
$C_2$: false 		& Increasing & 		& 		& Decreasing \\
$C_3$: false 		&		 & Decreasing & Increasing & 		  \\
$C_2$: false, $C_1$: true & & Increasing & Decreasing & 				  \\
$C_3$: false, $C_1$: true & Decreasing & 		&		 & Increasing \\
\hline
\end{tabular}
\caption{\label{tab1}Dynamical behaviour of a metric function in the proper time depending on the signs of the potential, the Hamiltonian and its initial and final values.}
\end{table}
\section{Bianchi class $A$ isotropisation in scalar-tensor theory} \label{s3}
The isotropisation of the Bianchi class $A$ models in General Relativity with a massive scalar field was already studied in several papers\cite{Fay03,Fay04}, using the ADM Hamiltonian formalism, when the Universe expands forever ($\Omega\rightarrow -\infty$). Defining the function $\ell$ of the scalar field: 
\begin{displaymath} 
\ell=\frac{\phi U_\phi}{U(3+2\omega)^{1/2}}
\end{displaymath} 
these papers rested upon the following assumptions:
\begin{enumerate}
\item The potential is positive. This is reasonable if we suppose that it could play the role of a varying cosmological constant.
\item $3+2\omega$ is positive, i.e. the scalar field weak energy condition is respected.
\item Near the isotropic state, the perfect fluid density parameter $\Omega_m$, the scalar field energy density parameter $\Omega_\phi$ and the shear parameters $\Sigma_\pm$ tend to some equilibrium values, being non vanishing for $\Omega_\phi$. This assumption physically defines what we called a class 1 isotropisation\footnote{There are three isotropisation classes\cite{Fay04A}. The class 2 is similar to the class 1 but with $\Omega_\phi\rightarrow 0$ near the isotropisation state. For the class 3, the density parameters $\Omega_\phi$ and/or $\Omega_m$ do not reach some equilibrium values and may oscillate\cite{FayLum03}.} and is in agreement with what we observe today. For a more precise mathematical definition, see for example \cite{Fay04A}. 
\item The isotropic state is reached sufficiently quickly. Physically it means that the variations of $\Sigma_\pm$, $\Omega_m$, $\Omega_\phi$ and $\ell$ in the neighbourhood of the isotropic state compared to their constant equilibrium values do not have any dynamical consequence on the asymptotical behaviour of the metric functions and potential. Among others, when in the neighbourhood of the isotropy $\ell^2$ tends to a constant $\ell_0^2$ with a small variation $\delta\ell^2$, one assumes that $\int\ell^2+\delta\ell^2 d\Omega\rightarrow \ell_0^2\Omega$, the effects of $\delta\ell^2$ thus being negligible. Anew, for a more precise mathematical definition, see 
\cite{Fay04A}.
\end{enumerate}
With these assumptions, we showed the following results concerning the isotropisation of the Bianchi class $A$ models:\\
\\
\emph{
Let us consider the General Relativity with a massive scalar field and the function $\ell$. The asymptotic behaviour of the scalar field near the isotropic state is given by the asymptotical solution of the following differential equation when $\Omega\rightarrow  -\infty$:
\begin{equation}\label{phiE}
\dot\phi=2\frac{\phi^2 U_\phi}{U(3+2\omega)}
\end{equation}
A necessary condition for a class 1 isotropisation is that $\ell^2$ tends to a constant such as $\ell^2<3$ for a flat model, $3$ being the spatial dimension, or $\ell^2<1$ for a model with curvature, $1$ being the spatial dimension minus $2$. If $\ell$ tends to a non vanishing constant, the metric functions tend to $t^{\ell^{-2}}$ and the potential disappears like $t^{-2}$. If $\ell$ tends to vanish, the Universe tends to a De Sitter model with a cosmological constant.}
\\
\\
In what follows, we will use these results to study the isotropisation of the Bianchi class $A$ models for a scalar-tensor theory, i.e. with a scalar field non minimally coupled to the curvature. The action of this theory can be deduced from the action (\ref{actionE}) by using a conformal transformation of the metric
\begin{displaymath} 
g_{\alpha\beta}=G^{-1}\bar g_{\alpha\beta} 
\end{displaymath}
where $G$ is the gravitation function depending on the scalar field. The barred quantities are the ones of the new frame known as the Brans-Dicke frame, the former one being called the Einstein frame. Using this metric, the action (\ref{actionE}) is rewritten in the form:
\begin{equation}\label{actionBD} 
S = \int \biggl[ G^{-1}\bar R- \bar\omega\phi^{-1} \phi_{, \mu} \phi^{, \mu} - \bar U \biggr ] \, \sqrt{-{}^{(4)}\bar g} \, d^4x 
\end{equation} 
with the following relations binding the quantities of the two frames: 
\begin{equation}\label{r1} 
dt=-Nd\Omega=\sqrt{G^{-1}}\bar dt
\end{equation} 
\begin{equation}\label{r2}
e^{-\Omega}=G^{-1/2}e^{-\bar\Omega} 
\end{equation} 
\begin{equation}\label{r3} 
U=\bar U G^2 
\end{equation}
\begin{equation}\label{r4} 
\frac{\omega + 3/2}{\phi^2}=3/2(G^{-1})_\phi^2G^{2}+\bar\omega  G\phi^{-1}
\end{equation} 
The necessary conditions for isotropisation of the Bianchi class $A$ models with a non minimally and massive scalar field remain identical to those found in the Einstein frame. The conformal transformation does not change the fact that the Universe isotropises: if it occurs in the Einstein frame, it will occur in the Brans-Dicke frame. To express these conditions in the Brans-Dicke frame, it is enough to replace $U$ and $\omega$ by their expressions (\ref{r3}-\ref{r4}) in the $\ell$ function. On the other hand the asymptotic behaviours of the metric functions and the potential can not be expressed any more in a closed form according to the proper time $\bar t$ because of the differential relation (\ref{r1}) which implies an integral of the gravitation function $G$.\\ 
The two following subsections present two methods to study the Bianchi class $A$ models isotropisation for the scalar-tensor theory specified by the action (\ref{actionBD}) on the basis of the above results.
 \subsection{Isotropisation of a scalar-tensor theory when the functions $G$, $\omega$ and $U$ are known}\label{s31}
\begin{figure}[h]
 \begin{center}
  \includegraphics[width=\textwidth]{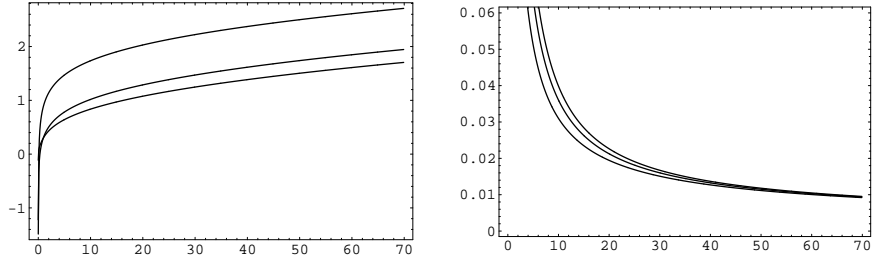}
 \end{center}
\caption{\label{resIso}\scriptsize{Logarithm of the metric functions (first graph) and their derivatives (second graph) with respect to the proper time and $\bar\omega_0=2.3$ and $k=-1$. Then $\ell^2=1.18$ and the Universe isotropises: the metric functions derivatives tend to a common behaviour,the metric functions growing at the same rate.}}
\end{figure}

\begin{figure}[h]
 \begin{center}
  \includegraphics[width=\textwidth]{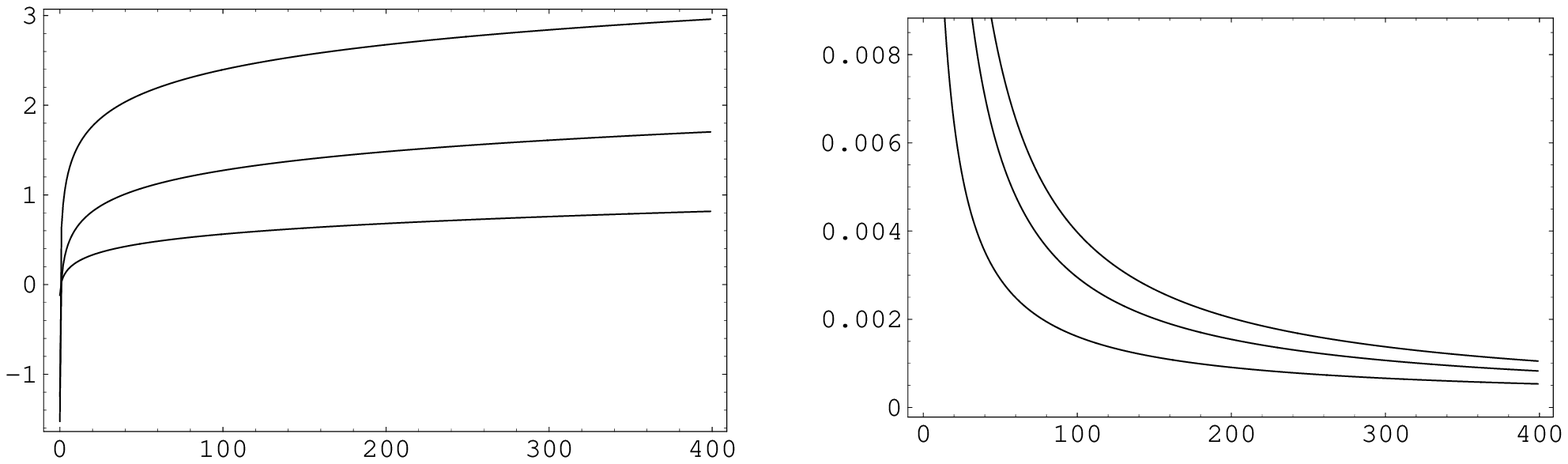}
 \end{center}
\caption{\label{resNoIso}\scriptsize{Logarithm of the metric functions (first graph) and their derivatives (second graph) with respect to the proper time and $\bar\omega_0=2.3$ and $k=-3$. Then $\ell^2=3.28$ and the Universe does not isotropise whatever the length of the time integration: the metric functions derivatives do not tend to a common behaviour and the metric functions deviate from each other.}}
\end{figure}

Supposing that one knows $G(\phi)$, $U(\phi)$ and $\omega(\phi)$, we will use the following algorithm to calculate the asymptotic behaviour of the metric functions when the Universe isotropises in presence of a non minimally coupled scalar field: 
\begin{enumerate} 
\item We calculate $\phi(\Omega)$ using (\ref{phiE}) what enables us to evaluate $\ell$ and thus the asymptotic form of $e^{-\Omega}$ according to the proper time $t$.
\item We thus get $\Omega(t)$,  $\phi(t)$ and then $\bar t(t)$ thanks to (\ref{r1}).
\item Using (\ref{r2}), we obtain $e^{-\bar\Omega}$.
\end{enumerate} 
Let us apply this method to the Brans-Dicke theory and the string theory at low energy with power and exponential potentials of the scalar field :\\
\begin{itemize}
 \item  \emph{Brans-Dicke theory with a potential in power of the scalar field}\\
 \end{itemize} 
This theory is defined by $G^{-1}=\phi$, $\bar\omega=\bar\omega_0$ and  $\bar U=\phi^k$. Consequently $U=\phi^{k-2}$ and $\ell$ is a constant being
\begin{displaymath}
\ell=\frac{k-2}{\sqrt{3+2\bar\omega_0}}  
\end{displaymath} 
If $k\not= 2$, we find that in the Einstein frame, when the Universe isotropises
\begin{displaymath}  
\phi\propto t^{2/(2-k)}  
\end{displaymath}
\begin{displaymath}  
e^{-\Omega}\propto t^{\frac{3+2\bar\omega_0}{(k-2)^2}}  
\end{displaymath}
with $k>2$ so that the scalar field is real. It follows from (\ref{r1}) that $t\propto \bar t^{(k-2)/(k-1)}$ and near the isotropic state, the scalar field and the metric functions behave in the Brans-Dicke frame like: 
\begin{displaymath}  
\phi\propto  \bar  t^{2/(1-k)}
\end{displaymath} 
\begin{displaymath}
e^{-\bar\Omega}\propto  \bar  t^{(1+k+2\bar\omega_0)/((k-2)(k-1))}
\end{displaymath}
\\ 
When $k=1$, one has $t\propto e^{\bar  t}$ and thus  $\phi\propto  e^{2\bar  t}$ and the Universe in the Brans-Dicke frame tends to a De Sitter model.\\
When $k=2$, $\ell$ is zero and the scalar field asymptotically tends to a constant. Consequently, $t\rightarrow \bar  t$ and the Universe tends to a De Sitter model in the Einstein and Brans-Dicke frames.\\
These results agree with those of  \cite{BilColIba99} and are illustrated on figures \ref{resIso} and \ref{resNoIso} for the Bianchi type $I$ model. The first one is such as $\ell^2<3$ and shows that the metric functions derivatives converge to a same behaviour whereas the metric functions grow at the same rate. The second one is such as $\ell^2>3$ and the Universe does not isotropise: the metric functions derivatives never converge to a common behaviour whatever $t$ and the metric functions deviate from each other. Some similar figures may be obtained for the three others scalar-tensor theories that we consider in this section.
\begin{itemize}  
\item  \emph{Brans-Dicke theory with an exponential potential}\\  
\end{itemize} 
This theory is defined by $G^{-1}=\phi$, $\bar\omega=\bar\omega_0$ and  $\bar U=e^{k\phi}$ and is thus such as $U=e^{k\phi}\phi^{-2}$. $\ell$ is then a function of the scalar field:
\begin{displaymath}  
\ell=\frac{k\phi-2}{\sqrt{3+2\bar\omega_0}}
\end{displaymath} 
In the Einstein frame, we calculate in the neighbourhood of the isotropic state that the scalar field behaves like: 
\begin{displaymath}  
\phi=2/(k-\phi_0e^{4\Omega/(3+2\omega_0)})
\end{displaymath} 
where  $\phi_0$ is an integration constant. It thus tends exponentially to the constant $2/k$ when $\Omega\rightarrow  -\infty$. Consequently $\ell$ tends to zero, $t\rightarrow \bar  t$ and the Universe tends to a De Sitter model in the Einstein and Brans-Dicke frames.\\
 \begin{itemize}  
 \item  \emph{String theory with an exponential potential}\\
 \end{itemize} 
This theory is defined by $G^{-1}=e^{-\phi}$, $\bar\omega=\phi  e^{-\phi}$ and  $\bar U=e^{k\phi}$. Consequently, $U=e^{(k+2)\phi}$ and it comes that 
\begin{displaymath}
 \ell=\frac{(2+k)\phi}{\sqrt{3+2\phi  e^{-\phi}}}
 \end{displaymath} 
In the Einstein frame and in the vicinity of the isotropy, we deduce from (\ref{phiE}) that 
\begin{displaymath}
 ExpIntegralEi(-\phi))-3/\phi=2(2+k)\Omega+\phi_0  
 \end{displaymath}
where $ExpIntegralEi(-\phi)=\int_\phi^\infty e^{-x}x^{-1}dx$. We deduce that if $k>-2$ (respectively $k<-2$), $\Omega\rightarrow -\infty$ when $\phi\rightarrow 0_+$ (respectively $\phi\rightarrow 0_-$). Then a power series of this equation about $\phi=0$, learns us that 
 \begin{displaymath}  
 \phi\rightarrow \frac{-3}{2(2+k)\Omega+\phi_0}  
 \end{displaymath} 
Consequently $\ell\rightarrow 0$ like $-3\left[2(2+k)\Omega\right]^{-1}$, and the Universe tends to a De Sitter model in the Einstein frame. We also find that $t\propto \bar  t$ and thus in the Brans-Dicke frame, $\phi\propto  t^{-1}\propto \bar t^{-1}$ and the Universe again tends to a De Sitter model. It is the same for the special value $k=-2$.
\\
 \begin{itemize}  
 \item  \emph{String theory with a power potential}\\
 \end{itemize} 
This theory is defined by $G^{-1}=e^{-\phi}$, $\bar\omega=\phi  e^{-\phi}$ and $\bar  U=\phi^k$ and thus $U=e^{2\phi}\phi^k$. It comes that
\begin{displaymath}  
\ell=\frac{k+2\phi}{\sqrt{3+2\phi e^{-\phi}}}  
\end{displaymath} 
and we asymptotically find that in the Einstein frame the scalar field is solution of 
\begin{displaymath}
 e^{k/2}ExpIntegralEi(-\frac{k}{2}-\phi)+\frac{3}{k}\ln\frac{\phi}{k+2\phi}=2\Omega+\phi_0
 \end{displaymath} 
Consequently  $\Omega\rightarrow  -\infty$ when $\phi\rightarrow 0$ with $k>0$ and a power series of this equation about $\phi=0$ shows us that 
\begin{displaymath}
 \phi\rightarrow  e^{\frac{k}{3}(2\Omega+\phi_0)}
\end{displaymath} 
It implies that near the isotropic state $\ell\rightarrow  k/\sqrt{3}$ and in the Einstein frame
\begin{displaymath}
\phi\rightarrow t^{-2k^{-1}}
\end{displaymath}
\begin{displaymath}
e^{-\Omega}\rightarrow t^{3 k^{-2}}
\end{displaymath}
Consequently $t\simeq \bar t$ and it comes in the Brans-Dicke frame
\begin{displaymath}
 \phi\propto  \bar  t^{-2k^{-1}}  
\end{displaymath} 
\begin{displaymath}  
e^{-\bar\Omega}\propto  \bar  t^{3k^{-2}}
\end{displaymath}
These results show the importance of the power or exponential law of the proper time as an asymptotic form for the metric functions in the vicinity of the isotropy in both Einstein and Brans-Dicke frames. In the following section, assuming these forms of funtions for $e^{-\bar\Omega}$, we examine the asymptotical behaviours of the gravitational function $G$ and potential $U$ of the scalar field when the Universe isotropises.
\subsection{Isotropisation of a scalar-tensor theory when $e^{-\bar\Omega}$ is a known function of the proper time}\label{s32}
The above theories show that when the Universe becomes isotropic, the metric functions of the Brans-Dicke frame asymptotically behave like some power or exponential functions of the proper time, typical of the solutions found without a scalar field or with a cosmological constant. In fact they are among the most usually widespread asymptotic behaviours in the literature (De Sitter model, FLRW models with a perfect fluid, etc). Reversing the steps of the previous section, we are going to determine the asymptotic forms of the potential and the gravitation function near the isotropic state when the metric functions of the Brans-Dicke frame tend to $\bar t^k$ or  $\bar e^{kt}$. We will thus get the conditions allowing to $G$ and $U$ to be bounded when the Universe isotropises, which currently seems to be the case, and will compare them with those allowing the isotropisation of the Universe. For this purpose, we will use the following algorithm:
\begin{enumerate}  
\item For each form of $\bar\Omega(t)$, we will suppose that $\ell$ tends to a vanishing or non vanishing constant what implies respectively that $e^{-\Omega}$ tends to $t^{\ell^{-2}}$ or to an exponential of $t$. 
\item We then calculate $G^{-1}(t, \bar  t)$ using (\ref{r2}).  
\item We deduce from it $\bar  t(t)$ using (\ref{r1}) and thus $G^{-1}(\bar t)$ and  $\bar  U(\bar  t)$ using (\ref{r3})
\end{enumerate} 
Employing this method, we obtain the following results:\\
\begin{itemize}
 \item When $e^{-\bar\Omega}\rightarrow \bar  t^k$ and the Universe isotropises such as $\ell\rightarrow cte\not = 0$, $G^{-1}\rightarrow \bar t^{2(1-k\ell^2)/(\ell^2-1)}$ and  $\bar  U=\bar t^{-2\left[(1+k)\ell^2-2\right]/(\ell^2-1)}$.  
 \item When $e^{-\bar \Omega}\rightarrow \bar  t^k$ and the Universe isotropises such as $\ell\rightarrow 0$, $G^{-1}\rightarrow \bar  t^{-2}$ and  $\bar  U\rightarrow \bar  t^{-4}$. It is in accordance with the previous point.
 \item When $e^{-\bar \Omega}\rightarrow  e^{k \bar  t}$ and the Universe isotropises such as $\ell\rightarrow  cte\not = 0$, $G^{-1}\rightarrow e^{2k\ell^2/(1-\ell^2)\bar  t}$ and  $\bar U\rightarrow  e^{2k\ell^2/(1-\ell^2)\bar  t}$.  
 \item When $e^{-\bar\Omega}\rightarrow e^{k  \bar  t}$ and the Universe isotropises such as $\ell\rightarrow 0$, $G^{-1}\rightarrow G^{-1}_0$ and  $\bar U\rightarrow  \bar  U_0$, $G^{-1}_0$ and  $\bar  U_0$ being two constants. It is in accordance with the previous point.\\
\end{itemize} 
These results are in agreement with those of the preceding section and generally stable as long as $e^{-\bar \Omega}$ and $e^{-\Omega}$ approach their asymptotic behaviours (in power or exponential of the proper time) more quickly than respectively $\bar  t^{-1}$ and $t^{-1}$. We discuss their significance in the following section.

\section{Discussion}\label{s4}
In this work, we explained in section \ref{s122} the main steps to obtain the ADM Hamiltonian of the General Relativity with a minimally coupled and massive scalar field. We then showed in section \ref{s2} how to use this formalism to find, for the homogeneous but anisotropic flat Bianchi type $I$ model, the dynamics of the metric functions according to the initial and final values of the Hamiltonian as well as the sign of the potential. We now comment these results. In what follows we will call "Big-Bang singularity" a pointlike singularity characterised by the vanishing of the three metric functions. A "pancake singularity" or "cigar singularity" will apply, respectively to the cases where one or two metric functions vanish.\\
From the table \ref{tab1}, we obtain the following results when the sign of the potential does not vary:
\begin{itemize}
\item A metric function can have a maximum (minimum) only in presence of a negative (positive) potential. Moreover, all the conjugate momenta $p_{ij}$ can not have the same sign and then the condition $C_1$ can not be true for all the metric functions: it follows that at least one metric function will not have any extremum.
\item Since $C_1$ can not be checked for all the metric functions, when the Hamiltonian is positive, the three metric functions can be increasing together at late times, but not decreasing. All types of singularity, Big-Bang type, pancake type or cigar type are possible at early time.
\item Since $C_1$ can not be checked for all the metric functions, when the Hamiltonian is negative, the three metric functions can be decreasing together at late time but not increasing. The singularity if it exists will only be of pancake or cigar type at early time.
\item From Collins and Hawking isotropy definition\cite{ColHaw73}, the Universe will isotropise when $\Omega\rightarrow -\infty$. It corresponds to the vanishing of $d\beta_\pm/dt \propto e^{3\Omega}$  and is necessary to the convergence of $(dg_{ij}/dt)/g_{ij}$ to a common value allowing to observe the same Hubble function on all the directions. We thus easily deduce, that Universe isotropisation will arise at late (early) $t$ times if the Hamiltonian is positive (negative).
\end{itemize}
As written in the first point, as long as the potential has a constant sign, the metric function can have one and only one extremum. This is also the case when we consider a flat or open FLRW model with trace-free matter, $\phi$ finite and $\omega_\phi>0$ as shown in \cite{BarPar97}. In this paper it is also proved that flat FLRW models can only contain a single minimum whereas here, we shown that a single maximum is also allowed for a negative potential. When the sign of the potential varies, the results of table \ref{tab1} are always true but $H_{ini}$ and $H_{end}$ define the intervals of values of the Hamiltonian for which the sign of the potential is constant.\\
\\
Using a conformal transformation of the metric whose rules are given by (\ref{r1}-\ref{r3}) and the results about the Universe isotropisation obtained within the framework of a minimally coupled and massive scalar field, we studied in section \ref{s3} the isotropisation of several scalar-tensor theories, spread in the literature, for the Bianchi class $A$ models. We obtained the following results in the Brans-Dicke frame:\\
\emph{
\begin{itemize} 
\item Brans-Dicke theory: 
\begin{itemize}
\item With a power potential of the scalar field $\bar U=\phi^k$, a necessary condition for the isotropisation will be that $(k-2)^2(3+2\bar\omega_0)^{-1}<3$ or $<1$ according to the absence or the presence of curvature. So that the scalar field is real, it is necessary that $k\geq 2$. Then when $k>2$, the metric functions tend to $\bar  t^{(1+k+2\bar\omega_0)/((k-2)(k-1))}$ and the Universe is expanding if $1+k+2\omega_0>0$. Moreover, this expansion will be accelerated if $(k-2)^2(3+2\bar\omega_0)^{-1}<1$. Hence in presence of curvature, the Universe expansion is always accelerated when it isotropises. The potential $\bar  U$ tends to vanish as $\bar t^{\frac{2k}{1-k}}$, thus being able to explain the smallness of the cosmological constant at late times. However, the gravitation function $G$ diverges as $\bar t^{\frac{2}{k-1}}$ in disagreement with its small observational value. If $k=1$, the Universe isotropises without condition to a de Sitter model, the gravitation function tends to vanish but the potential diverges. If $k=2$, the Universe isotropises without condition to a de Sitter model and $U$ and $G$ tend to some non vanishing constants.  
\item With an exponential potential of the scalar field  $\bar U=e^{k\phi}$, the Universe isotropises without condition to a De Sitter model and is expanding.  $\bar  U$ and $G$ tend to some non vanishing constants.  
\end{itemize}  
\item String theory at low energy:
\begin{itemize}  
\item With a power potential of the scalar field  $\bar  U=\phi^k$, a condition necessary to the  Universe isotropisation will be that $k^2<9$ or $<3$ according to the absence or the presence of curvature and $k>0$. The metric functions tend then to  $\bar  t^{3k^{-2}}$ and the expansion of the Universe is thus accelerated if $k^2<3$, which is always the case in presence of curvature. $G$ tends to a non vanishing constant and $\bar U$ vanishes like $\bar t^{-2}$, explaining the smallness of the cosmological constant at late times.  
\item With an exponential potential $\bar  U=e^{k\phi}$, a necessary condition for isotropisation will be that $k\geq -2$ ($k<-2$) if the scalar field is positive (negative). The metric functions tend then to a De Sitter model and $U$ and $G$ to some non vanishing constants.
\end{itemize}  
\end{itemize}} 
These applications clearly show, contrary to what occurs for a minimally coupled scalar field, that the Universe isotropisation does not lead systematically to an expanding Universe. This comes owing to the fact that the conformal transformation does not preserve the signs of the metric functions derivatives: an expanding Universe in the Einstein frame may be contracting in the Brans-Dicke frame. If these two theories are among the most studied in the literature, power or exponential laws of the proper time for the metric functions are also the most widespread. This is why it appeared significant to us to study them. Hence, the De Sitter model is often used to solve the flatness problem and lots of FLRW models solutions with a dust or a radiative fluid generally behave like powers of the proper time. In  \cite{NavSerAli99} where the asymptotic dynamics of the scalar-tensor theories with a perfect fluid are studied for the FLRW models, the late time solutions of the metric functions when they converge to General Relativity, give place to power or exponential laws of the proper time. It is also the case in \cite{Ame99} where one looks for scaling attractors likely to produce an accelerated expansion. It thus strongly justifies the study of these forms of metric functions as the outcome of isotropisation process. One obtains the following conditions in the Brans-Dicke frame ensuring the non divergence of the potential and the gravitation function when the Universe isotropises and $e^{-\bar\Omega}$ tends to $\bar t^k$ or $e^{k\bar t}$:\\
\\
 \emph{When the Universe isotropises, is expanding and that asymptotically the metric functions behave like a power of the proper time  $\bar  t^k$ ($k>0$), it is necessary and sufficient so that the potential and the gravitation function do not diverge asymptotically that
\begin{itemize}  
\item when $\ell^2<1$, $\frac{1}{k}<\ell^2<\frac{2}{1+k}$: thus $k>1$ and the Universe expansion is always accelerated.
\item when $\ell^2>1$, i.e. only for a flat Bianchi type $I$ model since isotropisation does not occurs for a model with curvature in such a case, $\frac{2}{1+k}<\ell^2<\frac{1}{k}$: thus $k<1$ and the Universe expansion is thus decelerated
\end{itemize} 
If the metric functions behave asymptotically like an exponential of the proper time $e^{k\bar t}$ $(k>0)$, the potential and the gravitation function do not diverge only if $\ell\rightarrow 0$.\\\\
}
This important result shows that \emph{isotropisation and finiteness of the potential and gravitation function only occur when the Universe tends to a De Sitter model with $\ell\rightarrow 0$ or when the metric functions tend to a power law of the proper time representing an accelerated Universe with a (vanishing or not) curvature or a decelerated flat Universe}. Hence, the expansion being accelerated today, this result is clearly in accordance with the presence of at least a small curvature in our present Universe.\\ 
One can also obtain some results about the No-Hair theorem which states that in presence of a cosmological constant, the Universe tends to a De Sitter model. With regard to General Relativity with a massive scalar field, we showed in \cite{Fay01} that this result could be generalised to any theories such as $\ell\rightarrow 0$ (when the assumption 4 of the section  \ref{s3} is respected) which thus include the particular case $U=constant$ for which  $\ell=0$ strictly. For the scalar-tensor theories, the fact that $\ell$ tends to zero is not sufficient any more to imply the convergence of the Universe to a De Sitter model because the asymptotic behaviour of the metric functions also depends on that of $\phi$. On one hand, if the potential $U$ is a constant, $\ell$ is exactly zero and the scalar field tends to a constant, implying that the Universe behaves like a De Sitter model. On the other hand, if the potential $\bar  U$ is a constant but that $\ell$ tends to a non vanishing constant, $G^{-1}\rightarrow t^2$ and $e^{-\Omega}\rightarrow t^{(2-\ell^2)\ell^{-2}}$. Consequently, \emph{the presence of a cosmological constant in the Brans-Dicke frame does not imply necessarily the convergence of the Universe to a De Sitter model contrary to what occurs in the Einstein frame}.\\
\\ 
All these results were established thanks to the ADM Hamiltonian formalism. It is a powerful tool but which unfortunately remains difficult to use for the Bianchi class $B$ models. It would thus be interesting to deal with this class of models by using the results of \cite{RyaWal84} in which the divergence theorem is redefined.
\section*{Acknowledgment}
I warmly thank Doctor Christian Scheen for the careful reading and correction of the section on the ADM Hamiltonian formalism.


\begin{thebibliography}{10}

\bibitem{MatGuzUre99}
Tonatiuh Matos, Francisco~S. Guzm{\'a}n, and L.~Arturo Une{\~n}a-L{\'o}pez.
\newblock Scalar field as dark matter in the {U}niverse.
\newblock {\em Class.Quant.Grav.}, 17:1707--1712, 1999.

\bibitem{Fay03A}
S.~Fay.
\newblock Scalar fields properties for flat galactic rotation curves.
\newblock {\em Astronomy and Astrophysics}, 413:799, 2004.

\bibitem{Fay03}
S.~Fay.
\newblock Isotropisation of {B}ianchi class {A} models with curvature for a
  minimally coupled scalar tensor theory.
\newblock {\em Class. Quantum Grav}, 20, 7, 2003.

\bibitem{Fay04}
S.~Fay.
\newblock Properties of homogeneous cosmologies in scalar tensor theories.
\newblock {\em Accepted for publication in {"}Progress in Field Theory
  Research{"}, Nova Science Publishers, Inc.}, 2004.

\bibitem{Mis69}
C.~W. Misner.
\newblock Mixmaster {U}niverse.
\newblock {\em Phys. Rev. Lett.}, 22:1071--1074, 1969.

\bibitem{Mis69A}
C.~W. Misner.
\newblock Quantum cosmology.
\newblock {\em Phys. Rev.}, 186, 5:1319--1327, 1969.

\bibitem{RyaWal84}
Michael~P. Ryan and Sergio~M. Waller.
\newblock On the {H}amiltonian formulation of class {B} {B}ianchi cosmological
  models.
\newblock {\em gr-qc/9709012 (unpublished)}, 1984.

\bibitem{Fay00A}
S.~Fay.
\newblock Hamiltonian study of the generalized scalar-tensor theory with
  potential in a {B}ianchi type {I} model.
\newblock {\em Class. Quantum Grav.}, 17:891--902, 2000.

\bibitem{Nar72}
Hidekazu Nariai.
\newblock Hamiltonian approach to the dynamics of {E}xpanding {H}omogeneous
  {U}niverse in the {B}rans-{D}icke cosmology.
\newblock {\em Prog. of Theo. Phys.}, 47,6:1824, 1972.

\bibitem{Fay04A}
S.~Fay.
\newblock Isotropisation of {B}ianchi class {A} models with a minimally coupled
  scalar field and a perfect fluid.
\newblock {\em Class. Quantum Grav.}, 21, 6:1609--1621, 2004.

\bibitem{FayLum03}
S.~Fay and J.~P. Luminet.
\newblock Isotropisation of flat homogeneous cosmologies in presence of
  minimally coupled massive scalar fields with a perfect fluid.
\newblock {\em Accepted for publication in Class. Quantum Grav.}, 2004.

\bibitem{BilColIba99}
Andrew Billyard, Alan Coley, and Jesus Ib{\'a}{\~n}ez.
\newblock On the asymptotic behaviour of cosmological models in scalar-tensor
  theories of gravity.
\newblock {\em Phys. Rev.}, D59:023507, 1999.

\bibitem{ColHaw73}
C.~B. Collins and S.~W. Hawking.
\newblock Why is the universe isotropic.
\newblock {\em Astrophys. J.}, 180:317--334, 1973.

\bibitem{BarPar97}
John~D. Barrow and Paul Parsons.
\newblock The behaviour of cosmological models with varying-{G}.
\newblock {\em Phys. Rev}, D55:1906--1936, 1997.

\bibitem{NavSerAli99}
Navarro A., Serna A., and Alimi~J. M.
\newblock Asymptotic and exact solutions of perfect-fluid scalar-tensor
  cosmologies.
\newblock {\em Phys. Rev. D}, 59:124015, 1999.

\bibitem{Ame99}
Luca Amandola.
\newblock {\em astro-ph/9904120}, 1999.

\bibitem{Fay01}
S.~Fay.
\newblock Isotropisation of {G}eneralised-{S}calar {T}ensor theory plus a
  massive scalar field in the {B}ianchi type {I} model.
\newblock {\em Class. Quantum Grav}, 18:2887--2894, 2001.

\end{thebibliography}
\end{document}